# Detector and Telescope Development for *ProtoEXIST* and Fine Beam Measurements of Spectral Response of CZT Detectors

J. Hong[a*], A. Copete[a], J. E. Grindlay[a], S. V. Vadawale[a],

W. W. Craig[b], F. Harrison[c], W. R. Cook[c] and N. Gehrels[d]

[a] Harvard Smithsonian Center for Astrophysics, 60 Garden St., Cambridge, MA 02138
[b] Lawrence Livermore National Laboratory, Livermore, CA 94550
[c] California Institute of Technology, Pasadena, CA 91125
[d] Goddard Space Flight Center, Greenbelt, MD 20771

## ABSTRACT

We outline our plan to develop *ProtoEXIST*, a balloon-borne prototype experiment for the Energetic X-ray Imaging Survey Telescope (EXIST) for the Black Hole Finder Probe. EXIST will consist of multiple wide-field hard X-ray coded-aperture telescopes. The current design of the EXIST mission employs two types of telescope systems: high energy telescopes (HETs) using CZT detectors, and low energy telescopes (LETs) using Si detectors. With *ProtoEXIST*, we will develop and demonstrate the technologies required for the EXIST HETs. As part of our development efforts, we also present recent laboratory measurements of the spectral response and efficiency variation of imaging CZT detectors on a fine scale (~0.5 mm). The preliminary results confirm the need for multi-pixel readouts and small inter-pixel gaps to achieve uniform spectral response and high detection efficiency across detectors.

## 1. INTRODUCTION

*ProtoEXIST* is a balloon-borne pathfinder for technology development for the Energetic X-ray Imaging Survey Telescope (EXIST), under study for the proposed Black Hole Finder Probe. EXIST will be a next generation wide-field hard X-ray survey mission, monitoring the full sky every orbit, and it was selected for a concept study under the NASA's Beyond Einstein program.[1] Figure 1 shows the current design of EXIST in orbit configuration.[2] EXIST employs two different types of telescope systems: high energy telescopes (HETs) and low energy telescopes (LETs). Both systems consist of multiple wide-field hard X-ray coded-aperture telescopes. The HETs use a large array of CZT detectors (~6 $m^2$) with tungsten masks (~24 $m^2$), covering 10 – 600 keV, and the LETs use Si strip detectors (~1.4$m^2$) with tungsten masks (~5.6 $m^2$), covering 5 – 30 keV.

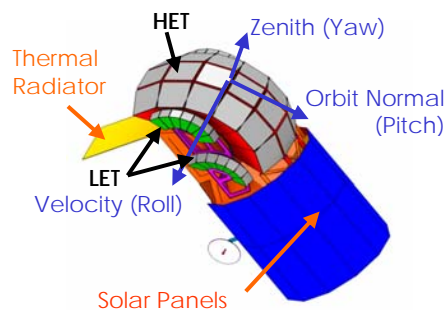

**Figure 1** Current concept for EXIST in orbit configuration.

The mission design will be further optimized in the future, but the overall design specs set by ambitious science goals will remain more or less similar. The diagram in Figure 2 summarizes the technology requirements for the EXIST HETs along with the corresponding mission parameters and driving science goals. For example, a large area of detectors (~6 $m^2$) is required to achieve the sensitivity goal and small pixels (1.25 mm) are necessary to resolve AGNs and galactic sources. These two factors drive the necessity of low-power ASICs for signal processing (~3×10$^6$ pixels) and an efficient detector packaging scheme. The wide energy bandwidth (10 – 600 keV) for the HETs will require relatively thick CZT detectors (≥ 5 mm), possibly with a depth sensing scheme to optimize imaging performance. In addition, an energy resolution of ~1– 3% is desired to identify the line emissions expected from some sources.

---

[*] Send correspondence to J. Hong (jaesub@head.cfa.harvard.edu)

The technologies we will explore using *ProtoEXIST* are specifically for the EXIST HETs. The EXIST HETs consist of a 3×6 array of identical sub-telescopes, each with an imaging array of ~3000 cm$^2$ of CZT detectors, ~1.2 m$^2$ tungsten masks, and CsI/Passive shields. Figure 3 shows a sub-telescope of the EXIST HETs and two different views of the EXIST instrument layout. The HETs are arranged to cover a contiguous ~65°×131° field of view (FoV, fully coded) around the zenith direction. In addition to hardware development for the EXIST HETs, *ProtoEXIST* will allow testing and demonstration of various aspects of the EXIST mission in a realistic space environment. For example, we will verify the performance of new imaging schemes we have developed for coded-aperture imaging with maximum sensitivity, such as radial hole imaging and scanning.[3,4] These new imaging schemes may be essential to achieve the EXIST science goals and some of the techniques can also be applied to the LETs. In section 2, we will review the *ProtoEXIST* concept and the development plan. As part of our current detector development efforts, we have devised a fine beam setup for spectral response and efficiency variation measurements of CZT detectors. In section 3, we will present the preliminary results of the measurements.

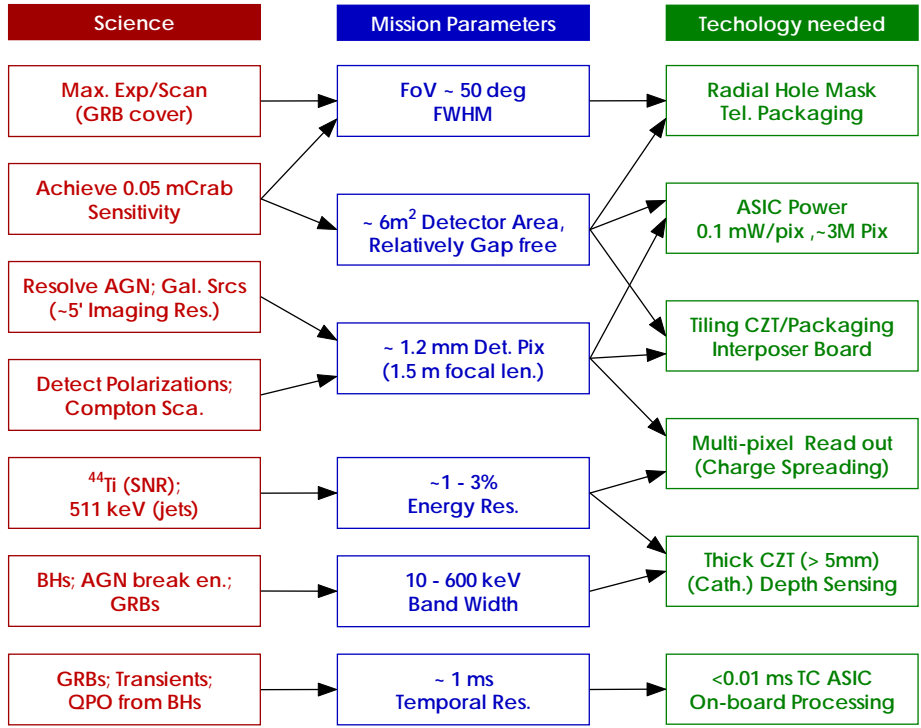
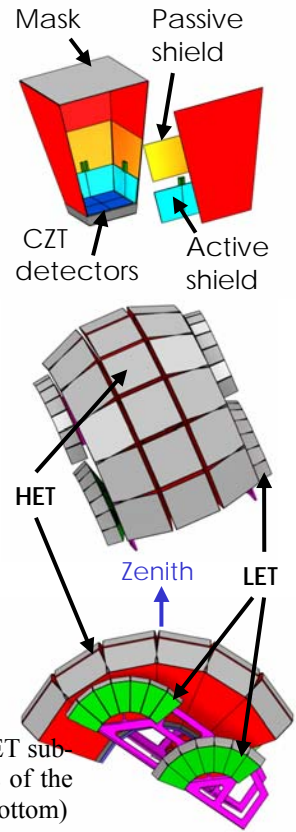

▲ **Figure 2** Technology requirements for the EXIST HETs along with corresponding mission parameters and science goals.

▶ **Figure 3** An exploded view of an HET sub-telescope (top) and two different views of the EXIST instrument layout (middle and bottom)

## 2. *PROTOEXIST*

One of the major elements for developing *ProtoEXIST* before EXIST is to develop technologies for the detector focal plane – large area imaging CZT detectors. To do so, we set a series of increasingly aggressive development plan – *ProtoEXIST1* and *ProtoEXIST2*. Table 1 compares the key mission parameters for the EXIST HETs, *ProtoEXIST1* and *ProtoEXIST*2. Under our proposed SR&T program, we plan to build a ~1024 cm$^2$ array of *ProtoEXIST1* CZT detectors and a ~256 cm$^2$ array of *ProtoEXIST2* CZT detectors. The detector size of *ProtoEXIST1* is moderately large enough to test practicality and complexity of the technologies required for a large-scale packaging scheme and to perform scientific observations using new imaging schemes. In the case of *ProtoEXIST2*, we will focus on developing small detector modules satisfying all the requirements (e.g. packaging and tiling of 1.25 mm pixel CZT arrays) for the detectors in the EXIST HETs.

Table 1. Key mission parameters of EXIST HET and *ProtoEXIST1 & 2*

| Parameters | EXIST HET | *ProtoEXIST1* | *ProtoEXIST2* |
|---|---|---|---|
| **Energy** | 10 – 600 keV | 30 – 600 keV | 30 – 600 keV |
| **No. of Modules** | 3×6 | 2×2 | 1[a] |
| **Mask (Tungsten)** | 5 mm thick, 2.5 mm pix | 5 mm thick, 5 mm pix. | 5 mm thick, 2.5 mm pix. |
| **Detector (CZT)** | 5 mm thick, 1.25 mm pix | 5 mm thick, 2.5 mm pix. | 5 mm thick, 1.25 mm pix. |
| **Area (Mod./Tot.)** | 56×56cm$^2$ / 5.6m$^2$ | 16×16cm$^2$/ 0.1 m$^2$ | 16×16cm$^2$ |
| **FoV (Mod./Tot.)[b]** | 21º ×21º / 65º ×131º | 9º ×9º /18º ×18º | 9º ×9º |
| **Ang. Res./5σ Loc.** | 5.7′/ 1.2′ | 17.2′/ 3.4′ | 8.6′/ 1.7′ |
| **Mask-Det. Sep.** | 1.5 m | 1.0 m | 1.0 m |
| **Temporal Res.** | < 1 ms | < 1 ms | < 1 ms |
| **Shields** | CsI/passive side, CsI rear shields | 1 CsI, 1 Plastic+Passive, 2 Passive (Pb-Sn-Cu) | 1 Active |
| **Sensitivity (5σ)** | 0.05 mCrab (<150 keV, ~1yr) 0.5 mCrab (>150 keV, ~1yr) | 5 mCrab (~3hr, < 100 keV) | 10 mCrab (~3hr, <100 keV) |

[a]Plan for the current SR&T. The complete version will have 2×2 modules. [b]Fully coded FoV.

### 2.1. Detector plane: Modularization for gap-free packaging of large arrays

The ability to build a semi-infinite detector plane without substantial gaps between units is a critical factor in constructing a large-area detector, given the limitations of the EXIST payload and to achieve optimal imaging performance under a coded-aperture system.[5] In order to explore this issue, we plan to adopt the following modularization scheme for *ProtoEXIST1*.

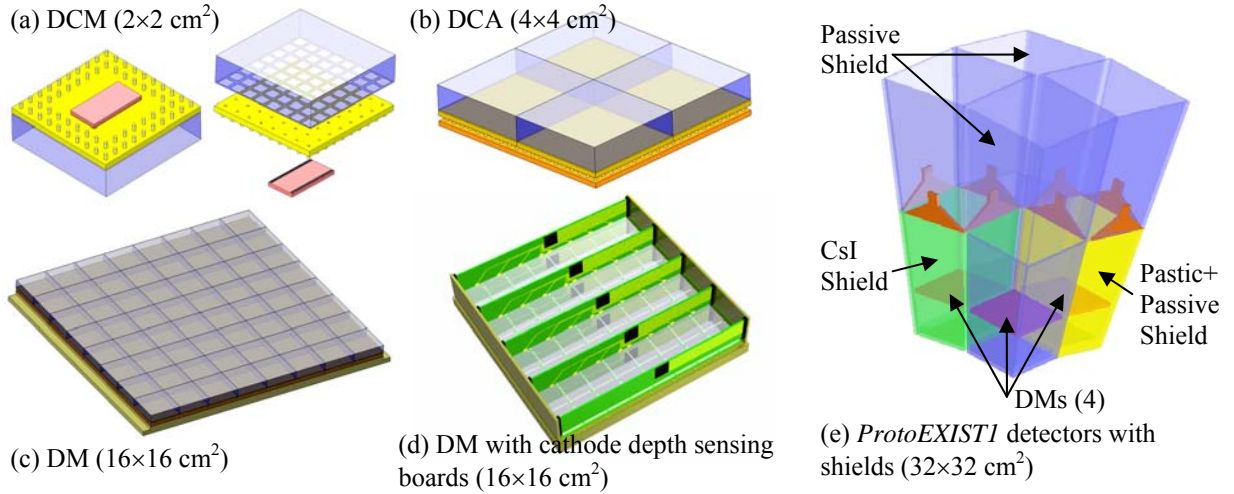

**Figure 4** Detector packaging scheme: (a) Detector Crystal Module (DCM) for single 2×2 cm$^2$ CZT and ASIC, (b) Detector Crystal Array (DCA), (c) Detector Module (DM) with (d) cathode depth sensing boards and (e) *ProtoEXIST1* detectors with shields.

First, we build a small detector unit called a detector crystal module (DCM, Figure 4a). A DCM consists of a 2x2 cm$^2$ CZT crystal bonded onto an interposer board (IPB), which traces the 8x8 array of anode pixel pads into a 1-D array of 64 input pads of an ASIC that handles the anode signals. The ASIC selected for *ProtoEXIST1* was developed for a Homeland Security application called RADNET.[6] We have selected the RADNET ASIC for *ProtoEXIST1* because of its approximate gain (~30 – 1000 keV), right form factor for 2.5 mm-pixel detectors, multi-pixel readout capability and low power consumption (100 μW/pixel, required for EXIST).

Second, we mount a 2x2 array of DCMs on the circuit board of an FPGA that controls and reads out the 4 ASICs, for a Detector Crystal Array (DCA, Figure 4b). The digitized pulse heights, pixel IDs and the time tag of each event in the 4 crystals will be processed by the FPGA. Third, we combine a 4x4 array of DCAs with a microprocessor to make a detector module (DM, 16x16 cm$^2$, Figure 4c). For flight tests of depth sensing, edge-on cathode readout boards will be mounted on one module (Figure 4d). Each cathode readout board will likely use an IDEAs VA32 ASIC (32 channels) to control cathode signals from 16 crystals along with 16 common mode channels.[3,7] Finally, *ProtoEXIST1* will have 2×2 independent DMs to complete a ~1024 cm$^2$ area of CZT detectors (Figure 4e). The modularization from DCM, DCA to DM is a key concept to achieve a virtually gap-free tiling of large area detectors for *ProtoEXIST* and EXIST. The actual number of units at each modularization stage may vary in the final EXIST design. For the current concept of the EXIST HETs, a DM consists of a 7×7 array of DCAs, and 2×2 arrays of DMs will comprise the detector plane for a sub-telescope (~3000 cm$^2$).

## 2.2. Small scale packaging: ASIC for *ProtoEXIST2*

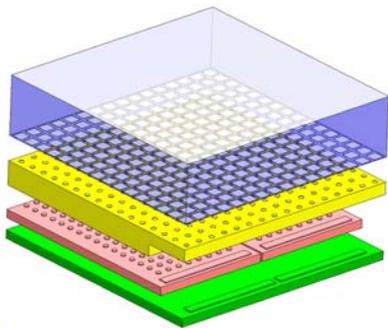

**Figure 5** DCM for *ProtoEXIST2* using a de-magnifying IPB (yellow) with 2 of 128 channel EXIST-type ASICs (red) mounted on FPGA board (green).

Many features in the RADNET ASIC are appropriate for the detectors in the EXIST HETs. However, the 1-D layout of input pads in the RADNET ASIC is not optimal for packaging 1.25mm pixel detectors. In collaboration with W. R. Cook (Caltech), we have started design studies of a new ASIC for the *ProtoEXIST2* detectors and ultimately for the EXIST HET detectors. Combining the long heritage of the ASIC development for the HEFT, RADNET, and STEREO projects,[6,8,9] the new ASIC will allow a proper packaging for the EXIST HET detectors by arranging the input pads in a 2-D array. A 2-D array of input pads allows a direct flip-chip bonding between an ASIC and a crystal. However, in order to achieve virtually gapless detector tiling, we will explore an option of employing IPBs that de-magnify the relatively large 2-D anode pixel arrays (1.25 mm pitch) into smaller 2-D input pad arrays (~ 1 mm pitch) of the new ASIC (Figure 5). The de-magnification allows room (~4 mm) to bring out ASIC control lines without introducing gaps between DCMs. To increase ASIC production yield, we will probably employ 2 128-channel ASICs for a DCM instead of a 256 channel ASIC. We will also explore the micro-via technology to bring control lines out of the ASIC efficiently.[10] The final layout of the new ASIC will depend on the feasibility and noise performance of the asymmetric de-magnifying IPB and the imaging performance under the presence of gaps between detector units.[5]

## 2.3. Shield and Mask

In addition to detector technologies, *ProtoEXIST* allows testing of various other aspects of the EXIST mission. For example, we will experiment with three different types of shielding schemes in order to find the optimal shield design for EXIST and to verify our background modeling capability. As shown in Figure 4e, each DM in *ProtoEXIST1* employs an independent shield scheme: active shields using ~1 cm/2 cm (side/rear) CsI scintillators for one DM, a combination of passive and plastic shields for another DM, and simple passive shields using Pb-Tn-Cu multi-layers for the other two DMs. In the case of active or plastic shields, the top portion of the side shields will be passive. The height of the side shields is about 50cm above the detector plane, and each DM will be mounted at an angle (9º) with respect to each other. The detector and mask distance is ~1 m, providing an 18º×18º fully coded FoV for *ProtoEXIST1*.

In the case of mask design, the wide energy band width (10 – 600 keV) and the fine angular resolution for the mission require a relatively large aspect ratio of mask elements (≥ 5 mm thick, 2.5 mm pixel pitch), which introduces self-collimation at large off-axis angles. We have introduced the radial hole imaging scheme to alleviate the self-collimation problem of relatively thick masks,[3,4] and we will verify the performance of radial hole masks using *ProtoEXIST*. The mask will be made of laser-etched multi-layer tungsten plates (5 layers, 1 mm thick each) with a varying pitch (on average, ~5 mm for *ProtoEXIST1* and ~2.5 mm for *ProtoEXIST2*) to orientate the holes in radial directions as done in our first prototype tests.[3] A URA pattern will be employed to optimize the imaging performance.

## 2.4. Telescope, Gondola, and Sensitivity

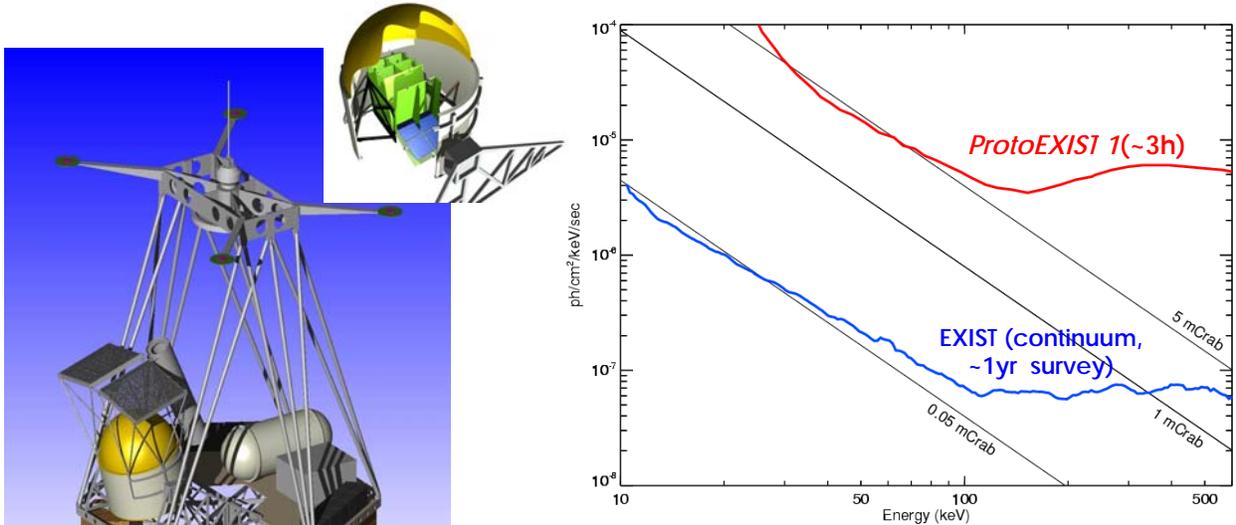

**Figure 6** The *ProtoEXIST* gondola (left), a cut-away view of the pressure vessel (middle inset) and the expected sensitivity of *ProtoEXIST* and EXIST (right).

Figure 6 shows the gondola of *ProtoEXIST* with masks, star camera and electronics systems and the inset shows an cut-away view of the pressure vessel, revealing the detector planes of *ProtoEXIST1 & 2*. The gondola design and construction techniques are based on the HEFT gondola. As for the pointing/aspect system, relatively coarse requirements for pointing and aspect accuracy (~1º pointing/ ~24″ aspect determination) do not impose a serious challenge. The scanning nature of the wide-field survey will also allow inertial pointing (or dithered scans). Scanning will be tested by rastering the FoV of *ProtoEXIST* slowly across a limited region of the sky, performing scanning coded-aperture imaging on a region of interest. Note that the *ProtoEXIST* telescopes are mounted in an offset elevation pivot that allows direct zenith observation without interference from the gondola structure. Finally, the expected sensitivity for 3-hour continuous observation of a source by *ProtoEXIST1*(assuming the source is always in the FoV) is ~5 mCrab at <100 keV. Figure 6 compares the *ProtoEXIST* sensitivity with the ~1-year survey sensitivity of EXIST (~0.05mCrab at < 150 keV). Note that for any given source in the sky, the 1-year survey by EXIST is equivalent to ~2.5-month continuous observation of the source (~20% duty cycle).

## 3. FINE BEAM MEASUREMENTS OF EFFICIENCY AND SPECTRAL RESPONSE OF CZT DETECTORS

We have been developing a series of measurement tools to calibrate and select a large number of CZT detectors appropriate for the project. These tools allow characterization of CZT crystals before permanently bonding to a detector system. Some of the results from the leakage current and radiation measurements have been reported earlier.[11] Here we report the preliminary results from a new addition to our tools, the fine beam measurement of efficiency and spectral response of CZT detectors. The main goal of this experiment is to measure the uniformity of the efficiency and the spectral response of an entire crystal on a very fine scale (< 0.4 – 0.5 mm). We want to explore these properties in relation to pixel pad (or gap) size, contact material, etc. This information can be used to find an optimal combination of crystals, contacts and readout schemes for highly uniform and sensitive CZT detectors.

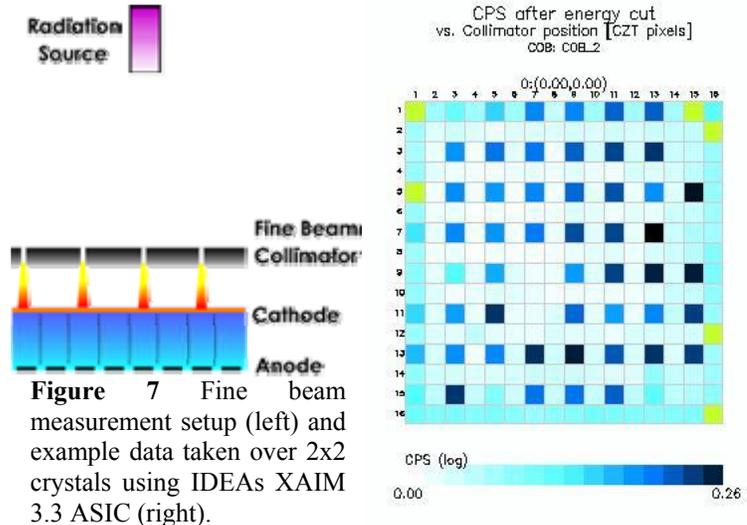

**Figure 7** Fine beam measurement setup (left) and example data taken over 2x2 crystals using IDEAs XAIM 3.3 ASIC (right).

### 3.1. Setup

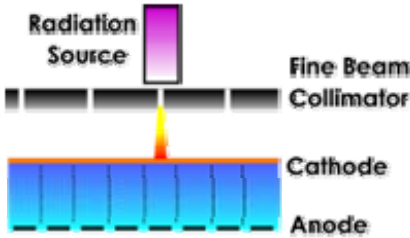

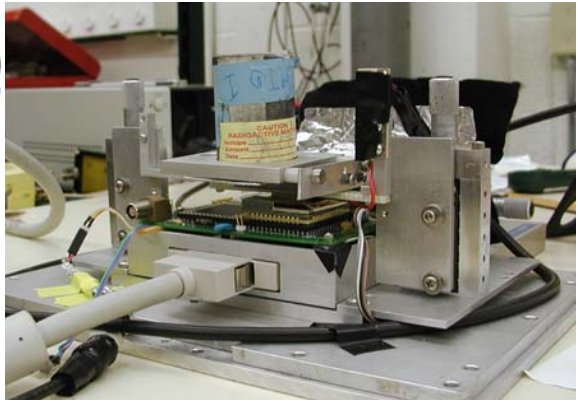

**Figure 8** Fine beam measurement setup exploring a small section of detectors (left) and the actual setup image (right)

The overall schematic of the setup is shown in Figure 7. The basic idea is to collimate the radiation using a collimator containing a series of small openings, so that the radiation on the cathode surface of the crystal is limited to a small area. Using ~0.1 mm-wide holes drilled through a 5 mm thick lead collimator, one can get ~0.4 – 0.5 mm diameter radiation beams on the cathode surface, which allows ~ 5×5 independent measurements for a 2.5 mm pixel. The holes are positioned over every other pixel in order to avoid the ambiguity in the beam locations arising from multi-pixel events due to charge spreading. The radiation source ($^{241}$Am) is placed relatively far away (>~30 inches) from the collimator to illuminate every hole uniformly in order to explore many pixels simultaneously. The $x$ and $y$ position of the collimator are precisely controlled by two micrometers to set the beam position accurately.

Figure 7 shows an example of such measurements over 2×2 IMARAD crystals (a DCA) with IDEAs XAIM 3.3 ASICs. The dark (blue) spots indicate the pixels that were radiated. The color intensity is set proportional to the count rate. There is a large variation in the count rate among the shined pixels. This non-uniformity is primarily due to the hole-to-hole variation of the opening geometry in the collimator (worsened by repetitive casual handling of the collimator, missing points in the checkerboard pattern indicate some holes are covered) and it is also partially due to the intrinsic efficiency variations in the detector. The setup is still being tested at this stage and the lead collimator will be replaced with a more robust multi-layer collimator – a lead layer sandwiched by Al or Cu plates for actual measurements. Meanwhile, we have performed a series of measurements focusing on a small section of the CZT detector by placing the radiation source near a single hole of the collimator, as illustrated in Figure 8. We will explore the results from these measurements.

### 3.2. Spectral Response of Inner Pixels in CZT detectors

First, we explore the boundaries of six inner pixels of a Pt/Pt contact IMARAD crystal (XAIM 3.3 ASIC for signal readout). Figure 9 shows the total count rates of the crystal from the $^{241}$Am source as a function of the beam centroids. The beam is ~0.4 – 0.5mm (diameter) wide. The pixel pad size is ~1.9×1.9 mm$^2$ with 2.46 mm pitch. Figure 9 shows the count rate maps in three different energy ranges: 30 – 70, 50 – 70 and 30 – 50 keV. In the left panel of Figure 9, the bright strips (one horizontal, two vertical) indicate the efficiency drop at the boundary of the pixels, revealing the explored region which contains 6 pixels. The two white square patches in the bottom middle pixel are the regions where we have skipped the measurement. A little skewing of the bright strips with respect to the *x-y* coordinates implies that the movement of the collimator is not perfectly aligned with the pixel pattern. The pixel gaps appear more clearly in the 50 – 70 keV range (the middle panel) because X-ray events occurring in the gap have a relatively larger chance to experience charge spreading, resulting in an incomplete energy deposition in one pixel. The XAIM 3.3 ASICs are not capable of multi-pixel readouts, and when charge spreading occurs over two or more pixels, the ASICs record only the peak pixel data. Therefore, when we select events in the 30 – 50 keV range, we predominantly select charge spreading events as shown in the right panel of Figure 9 (an almost reversed image compared to the middle panel). Figure 10 shows the same data as Figure 9, but it displays the count rate recorded only in the bottom middle pixel. In the 30 – 50 keV range (the right panel in Figure 10), we see a square ring pattern indicating the pixel pad and the surrounding gap.

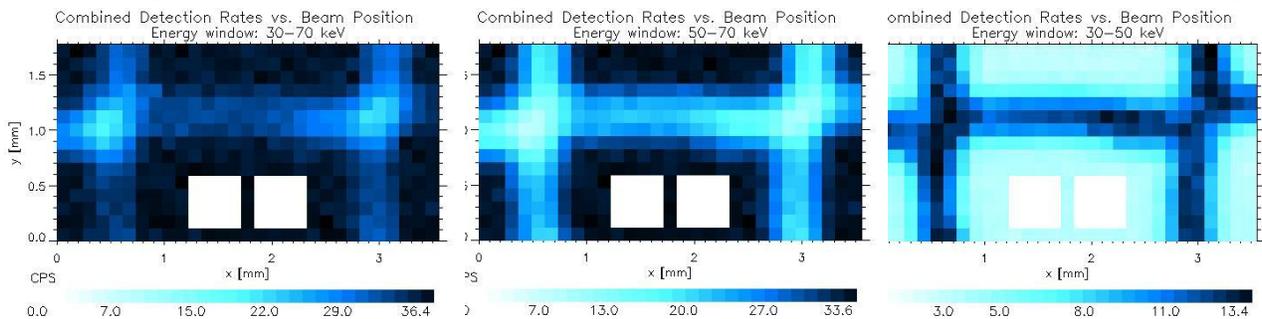

**Figure 9** Exploring boundaries of inner pixels: count rates vs. the beam centroids in 30 – 70 keV (left), 50 – 70 keV(middle), and 30 – 50 keV (right) range. We have skipped the two blank square regions in the bottom-middle pixel, expecting no systematic variation in the regions compared to the center of the pixel. The center of the bottom-middle pixel is at $x$=1.8 and $y$=0.0 mm.

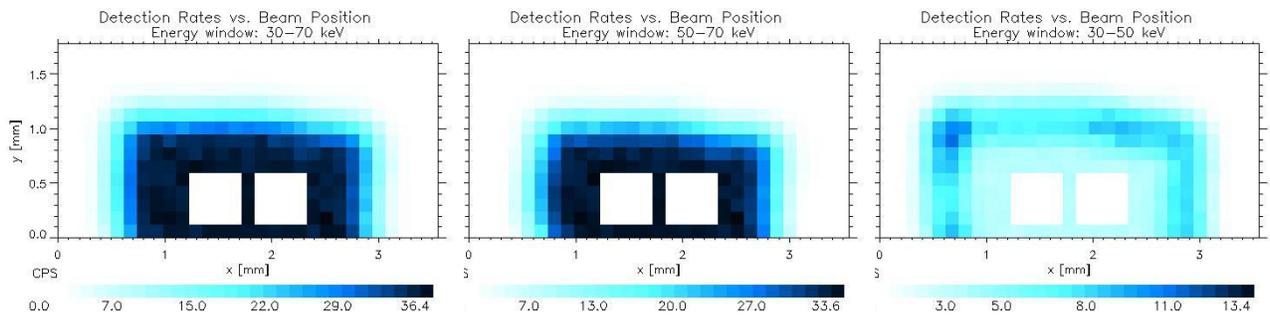

**Figure 10** The same as Figure 8 but showing the count rate recoded only in the bottom middle pixel.

Figure 11 summarizes the spectral response of the detector on various positions in a pixel. In the middle of the pixel, one can clearly see the 60 keV peak and even a lower energy line (dotted, red) from the $^{241}$Am source. At the boundary of the pixel, one can see a large fraction of events suffering charge spreading with the neighboring pixel (dashed, green). At the corner of the pixel, most events experience 3 or 4 way charge split among surrounding pixels, so that the spectrum does not feature the 60 keV peak from the $^{241}$Am source (dash-dotted, blue). These results strongly motivate multi-pixel readout schemes and narrow inter-pixel gap detectors for uniform spectral response and high detection efficiency.

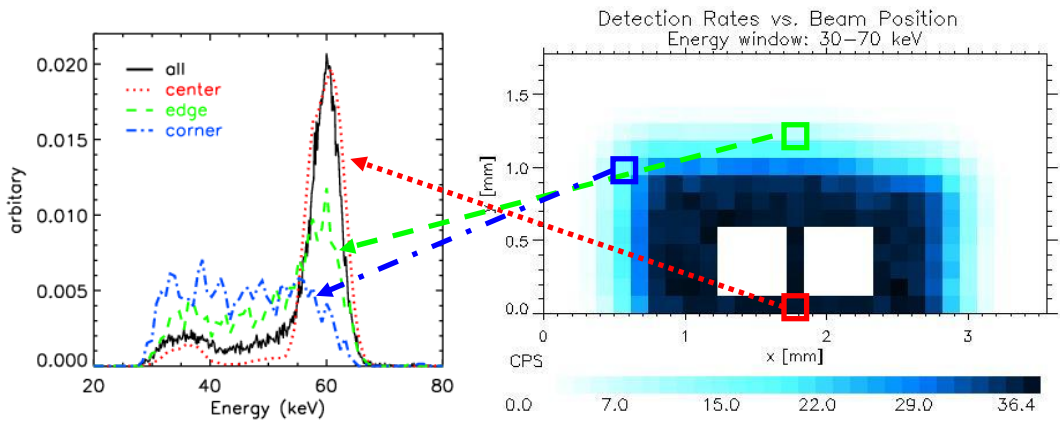

**Figure 11** Spectral Distributions vs. Beam Centroids in a pixel. The solid line (black) is for all the counts in the pixel, the dotted line (red) for events at the center of pixel, the dashed line (green) for events at the edge of the pixel, and the dash-dotted line (blue) for events at the corner of the pixel.

## 3.3. Edge pixels

Figure 12 shows another example, exploring the region of edge pixels (the detector edge is on the top side of the plots). The top two panels show the count rate in the 50 – 70 keV range, and bottom two panels in 30 – 50 keV. The left panels show the total count rate, and the right two panels are for count rates of an edge pixel. While the examples in Figure 10 do not show much variation among the six pixels, Figure 12 shows a serious deformation of detection efficiency at least in an edge pixel. There are many possible causes for this non-uniformity. For instance, the pixel pad on the detector might not have been placed properly during metallization, or during the bonding process with the circuit board, the conductive bonding material might not have been properly deposited. Interestingly, this particular crystal has an IMARAD guard band, which is a metallic strip on a plastic band that wraps around the side of the crystal. The guard band is used because it is known to lower the leakage currents of edge pixels. The edge pixel in the above example happens to be located right next to the gap of the guard band, and the gap may have altered the electric fields around the region. This will be studied further. Further measurements on various types of crystals and pixels are underway in order to find out the root cause of non-uniformities and ways to avoid them. In the above example, the fine beam measurement setup has been used for the XAIM ASIC based detectors, and the lack of multi-pixel readout capability in the XAIM

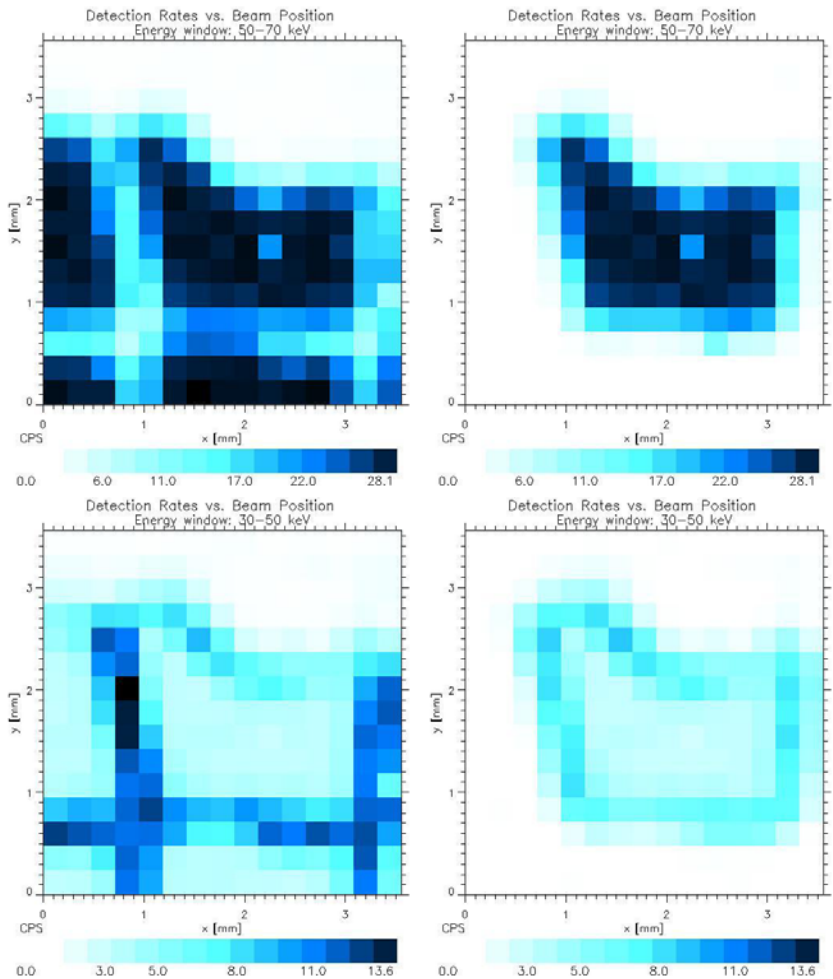

**Figure 12** Count rates vs. beam centroids around the edge pixels. The left panels for the total count rates, and the right panels for the count rates only in a pixel. The top two panels are for 50 – 70 keV and the bottom two panels for 30 – 50 keV.

ASIC limits the detailed study of multi-pixel events due to charge spreading or Compton events. The setup can be easily converted into the new system using RADNET ASICs that are capable of multi-pixel readout when the new detectors are ready. Such measurements will lead to very interesting results for optimizing detector performance regarding charge spreading and Compton events.

## 4. CONCLUSION

We have described our plan to develop *ProtoEXIST*, which will be a pathfinder for technology development for the proposed EXIST concept for the Black Hole Finder Probe. The plan focuses on extensive detector development that can lead a very large area of virtually gap-free CZT detector arrays. In addition to detector technology, *ProtoEXIST* will demonstrate various other aspects of the EXIST mission, from shield design/background modeling to new imaging schemes such as radial hole imaging and scanning. As a part of the on-going detector development, we have assembled various measurement tools to characterize the properties of CZT detectors. We have presented the initial results of new fine beam measurements, which motivate the necessity of multi-pixel readout schemes and small pixel gap sizes for highly efficient detectors with uniform spectral response.

## ACKNOWLEDGMENTS

This work is supported in part by NASA grant NAG5-5279 and NAG5-5396.